# MICROSYSTEM PRODUCT DEVELOPMENT

*Marc A. Polosky and Ernest J. Garcia*

Sandia National Laboratories
Albuquerque, New Mexico, USA 87185-0329

## ABSTRACT

Over the last decade the successful design and fabrication of complex MEMS (MicroElectroMechanical Systems), optical circuits and ASICs have been demonstrated. Packaging and integration processes have lagged behind MEMS research but are rapidly maturing. As packaging processes evolve, a new challenge presents itself, microsystem product development.

Product development entails the maturation of the design and all the processes needed to successfully produce a product. Elements such as tooling design, fixtures, gages, testers, inspection, work instructions, process planning, etc., are often overlooked as MEMS engineers concentrate on design, fabrication and packaging processes. Thorough, up-front planning of product development efforts is crucial to the success of any project.

## 1. INTRODUCTION

MEMS technology became popular around the beginning of the 90s'. Since then, many complex MEMS designs have been realized [1], [2]—see Figure 1. MEMS-based sensors and mechanisms have been produced and sold on the market; but, estimates of the numbers of new products introduced have been less than experts have predicted.

At the turn of the century, those skilled in the art of microsystems began to discover the lack of adequate packaging solutions. Much work has since transpired in developing and maturing MEMS packaging solutions. As packaging solutions evolve, the numbers of MEMS-based products are predicted to increase. Complete and thorough microsystem product development will be critical to the success of these endeavors.

There are three distinct phases commonly used to describe a product's history: 1) research and development, 2) product development, and 3) production.

Figure 2 depicts the flow from one to the next and the activities associated with each phase.

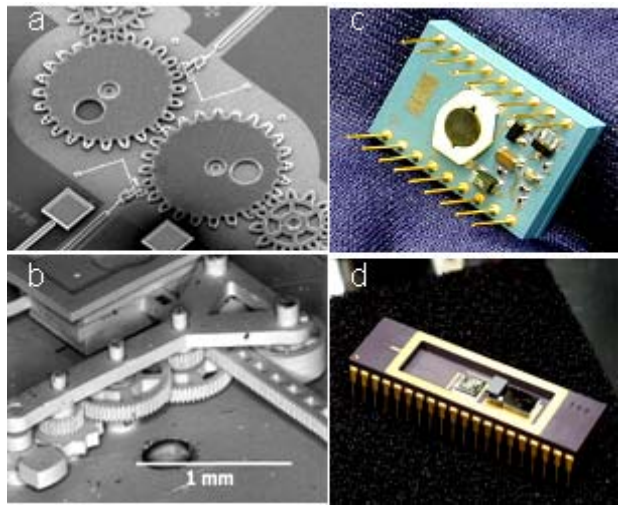

*Figure 1. Components fabricated in Sandia National Laboratories MEMS facilities a) MEMS micro gearing, b) LIGA fabricated mechanical regulator, c) Packaged micro optical switch, d) Packaged MOEMS device.*

Microsystem product development is that activity where an immature technology is developed sufficiently to move it from a low technology readiness level to one suitable for production. This has often been referred to as crossing the wall between research and product, crossing the valley of death or closing the technology readiness level gap [3]. This task is extremely difficult and at the same time is not viewed as a particularly appealing or glamorous task. Part of the difficulty lies in the fact this task is very often underestimated by management and staff.

A typical product development cycle is started with a schedule that is too tight, and with inadequate resources and staff available from the start. In addition, the requirements are generally poorly understood or ambiguous, and the true operating environments are unknown. Add to these, uncertainties in material behavior, including their time dependant behavior, and a lack of knowledge of the underlying physics and we have a recipe for failure

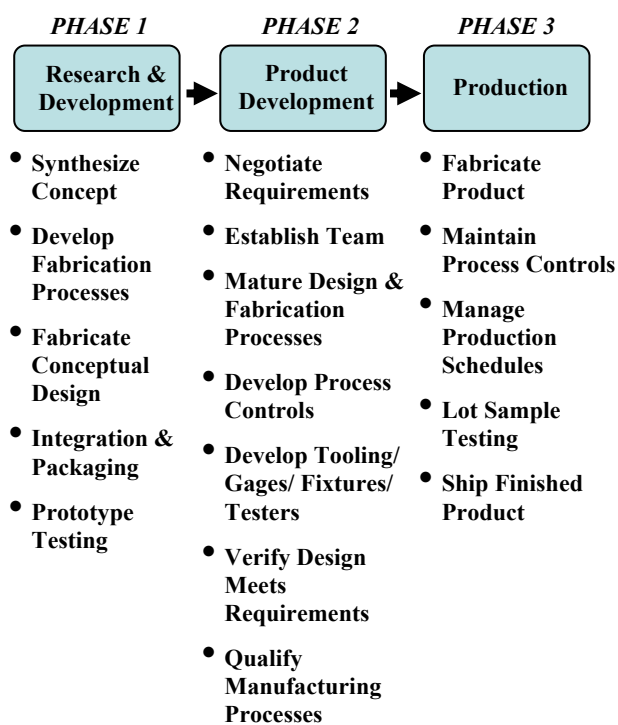

**PHASE 1**    **PHASE 2**    **PHASE 3**

Research & Development → Product Development → Production

- **Synthesize Concept**
- **Develop Fabrication Processes**
- **Fabricate Conceptual Design**
- **Integration & Packaging**
- **Prototype Testing**

- **Negotiate Requirements**
- **Establish Team**
- **Mature Design & Fabrication Processes**
- **Develop Process Controls**
- **Develop Tooling/ Gages/ Fixtures/ Testers**
- **Verify Design Meets Requirements**
- **Qualify Manufacturing Processes**

- **Fabricate Product**
- **Maintain Process Controls**
- **Manage Production Schedules**
- **Lot Sample Testing**
- **Ship Finished Product**

*Figure 2. Three phases of product manufacturing.*

Many MEMS presentations have been assembled focusing on the research and development but few have discussed product development and production. It has been the author's experience that product development is the most arduous phase of the three. The work entailed is not as glamorous as inventing new devices and fabrication processes. This paper identifies some of the tasks required to successfully develop a MEMS product.

## 2. MICROSYSTEM PRODUCT DEVELOPMENT

Product development begins once the research and development phase has been completed. At this juncture the supplier and potential customer have jointly reviewed the prototype test results, conducted a market analysis and depending on these results, concluded that beginning a product development phase is warranted, by customer demand and economic and technical viability [4]. The following subsections describe the product development work needed to mature a device for production.

### 2.1 Programmatic

The three primary constraints to any program are cost, schedule and performance [5]. All three must be thoroughly defined at the onset of the project. Many projects fail due to poorly defined, poorly understood, or ambiguous requirements. It is crucial to success to negotiate with the customer clear attainable device requirements at the onset. Unfortunately, it is not until late into the product development cycle, and after the expenditure of substantial capital, that device performance issues are discerned. Clearly this is unacceptable and fosters a poor relationship with the customer. MEMS designers must be capable of accurately predicting device performance before accepting customer requirements.

We believe it is good practice to negotiate at least three product development builds prior to start of production. The actual number depends on the particulars of the product being developed. This provides both the component designer and fabrication engineers the time needed to completely develop the device design and fabrication processes.

Six steps define the product development build cycle, they are design refinement, fabrication, integration, test, post mortem/failure analysis and requirements review— Table 1 describes each step and Figure 3 depicts the cycle. To mitigate development issues, the product development team works these steps concurrently. The process is commonly denoted concurrent engineering. We strongly support concurrent engineering for product development. The process promotes teaming on all aspects of development and enables timely responses for problem solving.

*Table 1. Product Development Steps*

| Product Development Step | Description |
|---|---|
| Design | Refine Device Design & Fabrication Processes |
| Fabrication | Process Development and Component Fabrication |
| Integration | Assembly Process Development |
| Test | Functional, Environmental, Reliability and Aging |
| Post Mortem/Failure Analysis | Failure Mechanisms and Physics |
| Requirements Review | Examine Performance, Cost, and Schedule |

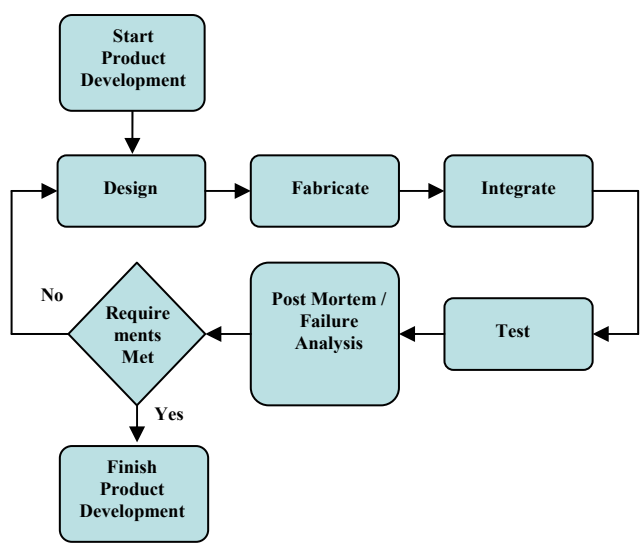

*Figure 3.   Recommended product development build cycle.*

Schedules should be established so there is sufficient time to complete each product development step. Often programs fall behind and an overlap between develop build cycles occur. As a result, design refinement for the subsequent step begins before sufficient time for testing and post mortem analysis is allowed. This defeats the intent of the multiple build cycles and creates opportunities for failure. Proper scheduling is imperative to avoiding cycle overlap.

Once device requirements and schedule are determined, an estimate of the overall project costs will be required. If the costs are too high, the customer and design team will need to make compromises to establish a balanced set of cost, performance and schedule. It is important for the design team to understand the time and effort required to develop the product and to not underestimate the task.

## 2.2  Design

The design phase entails refining the conceptual design to meet customer requirements. Accurate modeling of device performance is a must. This task will save the project time and money by selecting optimum device geometry to best meet performance requirements. Many software packages are available to help the designer with this effort.

Device models require accurate input parameters and valid assumptions. Often values are chosen based on information found in handbooks or from prior work. To ensure device models are accurate, test results from hardware builds must be compared with modeling results.

Where inconsistencies are identified, model parameters and assumptions should be modified to improve accuracy. This step is crucial to iterating towards the best design.

It is imperative the design team work closely with the fabrication team during the design phase of product development. Manufacturing limitations should be clearly understood by the design team while device performance requirements should be understood by the fabrication team. Understanding manufacturing limitations and performance requirements will reduce stressful interactions between team members and help focus the team on the work at hand. Each team will need to compromise in order to be successful.

A significant part of the product development phase is related to the packaging of a microsystem. Ideally this aspect was thoroughly considered during the R&D phase. However, in the usual case most of the resources and effort were spent developing the microsystem sub-elements. A successful product development can only conclude with the development of a packaging solution that meets all functional and environmental requirements at a cost-competitive price. As an example, a decision to package devices as individual die or to package as a large-scale waver-level process will have a significant impact on the final product in terms of cost, reliability and manufacturing complexity. Each possesses advantages and disadvantages which should be thoroughly examined before the choice is made.

Wafer-level packaging solutions are currently immature. Many issues need to be worked out to realize this option. The ability to get information or energy into and out the package can be challenging. Wafer bonding processes consistent with MEMS fabrication processes require development. If the package requires more than one wafer bonding step, difficulties begin to increase. On the plus side, wafer-level packaging reduces handling and contamination issues, problems that plague die level packaging.

If die-level packaging is selected, a decision will need to be made as to the use of custom or commercially available packages. Custom packages offer benefits of tailoring the package to the device needs. Figure 4 depicts a MOEMS device packaged in LTCC. Two optics chips, one MEMS chip and an ASIC are assembled in a Low Temperature Co-fired Ceramic (LTCC) custom package.

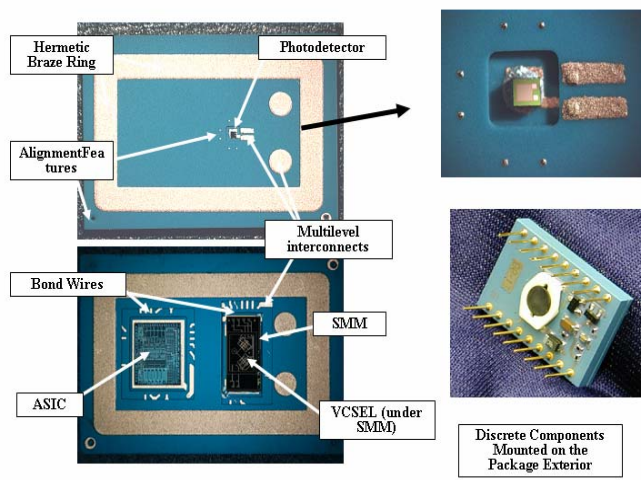

*Figure 4. MOEMS device packaged in custom LTCC package.*

If a custom package solution is selected, the design should be developed such that the device can be used with commercially available test sockets or connections. This will simplify testing.

The use of commercial packages can save the project time and money and should be considered as the first die-level packaging option. Prior work packaging ICs into commercial packages can be useful for MEMS projects [6].

During the selection of materials used for the device, care should be taken to ensure material compatibility. For devices with electrical contacts, the use of organics should be minimized. Out gassing degrades contact resistance over time. Thermal expansion coefficients should be matched for devices that will be exposed to large thermal gradients. The designer should consult experts in the field of materials to ensure proper materials selection.

The completion of the design phase should yield a complete set of drawings with tolerances. The drawing set will be used by the team to fabricate, assembly, and test the completed product. Care should be taken to control the drawing set so the correct version of the design is used throughout the development process. The drawing set should be available to the entire development team.

## 2.3 Fabrication

Successful product development engineering requires the development of a stable manufacturing process which minimizes process variation. Such a process will prevent defects from appearing in the final product. While a repeatable manufacturing process is the goal, it must be balanced with manufacturing costs since the cost of manufacture goes up as process variability is reduced. However there are many instances where process variability can be reduced by careful selection of process parameters via an optimization process. Product design engineers and manufacturing process engineers must work in a concurrent engineering environment which is a proven methodology for product development where tasks are conducted in parallel with early consideration of all segments of the product development cycle such as the product design segment and the manufacturing development segment.

Before manufacturing can begin, materials must be procured. As these materials arrive at the manufacturing plant, they must be inspected for purity and uniformity. For example, SOI wafers should be inspected for film thickness across the wafer. Buried oxide layer thickness should be measured and recorded. Securing a good manufacturer of raw materials is absolutely necessary since the underlying materials are critical to controlling device performance.

The manufacturing processes must be repeatable and controllable; this is essential. The use of process flow maps to identify all the manufacturing steps is a useful method to document and plan fabrication activities. The process flow map should include the specific details of each manufacturing step, i.e. deposition times, rates, temperatures, materials used, etc. A detailed process map helps assure every step is followed in the proper sequence and at the proper levels.

Manufacturing equipment must be routinely calibrated and maintained. Tool malfunction or breakdown is a common issue during product development and production. Where practical, for critical equipment, an inventory of spare parts should be maintained, especially parts that frequently need replacement and require long lead times to procure. This practice mitigates long schedule slips.

Control charts should be used for each manufacturing tool to monitor performance. These charts should be reviewed often to ensure the processing falls between acceptable limits. Processed wafers should be inspected to ensure the geometry is within limits and etch profiles are acceptable. Controlling manufacturing processes is crucial to ensuring device repeatability.

Device tracking is another important element. Tracking entails documenting the manufacturing equipment used to make the device, as well as the processing details and sensor feedback data acquired during manufacturing. The data is stored and later compared to device test results. Production engineers use the information to refine manufacturing processes.

The completion of the package phase should include segregated dice ready for integration. Care should be taken to store these parts in a clean and dry environment prior to delivery to assembly. Thought should be given to careful packaging for transportation to mitigate damage to the product.

## 2.4 Integration

Integration entails assembling the piece parts required to accomplish the product. For MEMS, this task can be quite challenging. MEMS integration requires handling, placement, bonding, interconnecting, sealing and marking. In many cases new tooling, gages and fixtures must be specially designed and fabricated. Tools developed for Integrated Circuits (ICs) often times must be adapted for MEMS integration work.

For die-level integration into packages, several issues are likely to be encountered: particle contamination, handling errors, inadequate die attachment, improper alignment, poor wire bonding, and incomplete sealing. Processing steps must be methodically developed to mitigate these common issues.

Prior to the start of integration, thorough and complete work instructions must be generated. These steps are used to instruct and train assembly technicians. Specifications related to cleaning, inspection, handling, attachment, etc., should be generated by the team prior to the start of production. Work instructions detail every stage in assembly to ensure steps are followed in the appropriate order.

The use of known good piece parts for assembly is paramount to achieving high yield. In the case of MEMS die, this can be difficult. Figure 5 shows a custom designed inspection fixture, also used for transportation, used to inspect MEMS acceleration switches. The fixture is loaded with dice and the backside is fitted to a vacuum tool. The moving elements on the MEMS die are pulled down by vacuum forces into the bottom cavity during inspection. By examination through a microscope, integration engineers can determine whether or not each die is suitable for assembly. The use of known good die reduces scrape and saves time and capital.

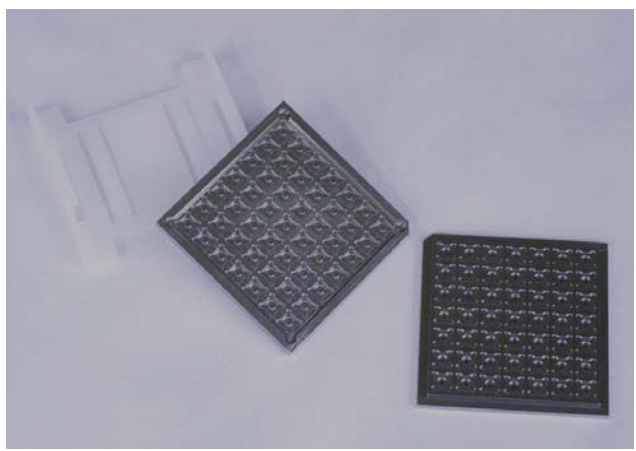

*Figure 5. Custom designed stainless steel waffle pack used for individual die inspection and as a transportation fixture.*

Integration engineers should keep an accurate count of piece parts. Long lead items should be ordered in advance so that they are on hand to meet schedule demand.

An accurate Record of Assembly (ROA) should be maintained so that information can be used in the event of testing failures. The ROA should be linked to the processing data base. Thus, integration and processing information are available to aid in failure analysis. This information is invaluable to the program.

## 2.5 Testing

To verify the device meets customer requirements testing is required. Four types of tests are commonly used to characterize and qualify product: in-process, environmental and functional, margin, and aging. In-process tests are conducted during the integration phase to ensure piece parts are functioning properly prior to sealing the device. These tests often offer the integration engineer the opportunity to rework the product.

Environmental testing entails testing the component to the normal operating conditions. These tests may include thermal, vibration, mechanical shock, pressure, electrical, etc. The results from these tests are used to validate the device meets customer requirements.

Margin testing or sometimes denoted over-testing, is used to determine device margin. These test results are used to characterize the robustness of the device. Customers are always interested in device margin.

Aging tests are used to determine the life of the component. They can include accelerated aging studies typically conducted in a thermal chamber where the device is subjected to many high and low thermal cycles and then tested for functionality. Another form of aging tests is life cycle tests. Here the unit is cycled until failure. These results are crucial to understanding the device performance.

Prior to the start of testing, testers, fixtures, and software code must be generated. This work should be started early and accounted for carefully in the product schedule. Often the time needed to complete fabrication of testers is under estimated and results in a slip in the schedule. Figure 6 depicts a custom fixture designed to test multiple MEMS acceleration switches in a centrifuge.

Post testing, an assessment will be needed to determine whether or not the product satisfies requirements. The generation of a requirements matrix is a useful tool for qualifying the product. The matrix documents the test and results and maps the information to the pertinent customer defined requirement. The tool is useful for device qualification.

Where the product fails to meet requirements, design, fabrication and integration processes will need to be modified to correct deficiencies. The product team should schedule sufficient time to review test results prior to the start of the next design refinement step. Test results, design geometry, processing variations, and record of assembly information will be useful for design and manufacturing refinement.

Device reliability will be determined by reviewing the test results and conducting a statistically based analysis. Each development group build should be thoroughly tested, the data analyzed, and a design review conducted. The review should address device performance, device produciblitiy and integration processes.

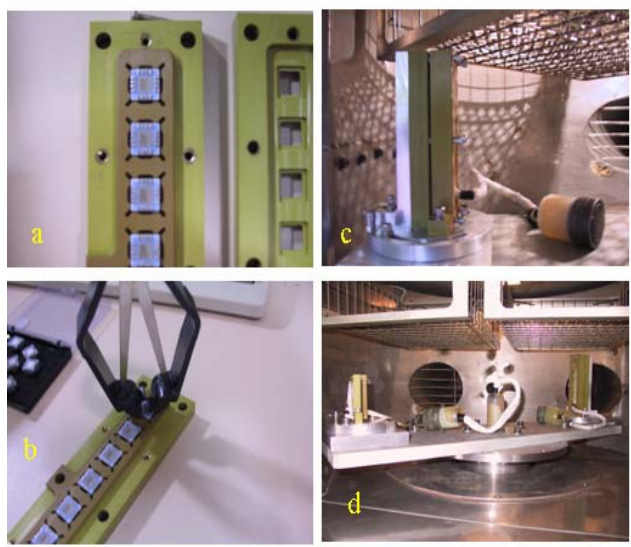

*Figure 6. Test fixture and tester used to qualify MEMS-based inertial switches.*

### 2.6 Post Mortem and Failure Analysis

Post mortem is an important but often overlooked product development step. This work takes place post testing and involves disassembly of the product for analysis. Lot samples are pulled from populations of product that have undergone environmental, margin and aging tests. The product is disassembled and thoroughly inspected for wear, fracture, contamination, debris, fatigue, etc... Tools such as time-of-flight SIMS for chemical analysis and SEM for visual and metrology are useful for discerning information. The results should be documented and reviewed by the development team prior to the next design refinement cycle. Well taken photographs are invaluable.

Failure analysis is a sub set of post mortem analysis. This is an often difficult task when intermittent device failures are encountered. Care must be taken during disassembly. It is easy to eliminate or damage evidence of the failure mode and thus limit the accurate discernment of the root cause of failure. The authors recommend the generation of a thorough well thought out failure analysis plan prior to the start of the post mortem. Figure 7 is a photograph taken during a failure analysis of a MEMS-based electrical switch. Silicon contamination was detected on mating 100 micron diameter contact surface thus rendering an open circuit condition.

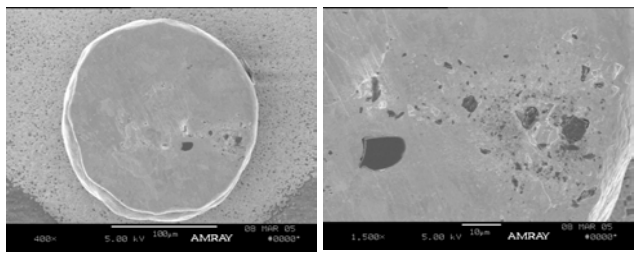

*Figure 7   Photograph of an electrical contact with particulate contamination captured during a failure analysis.*

## 3.0 CONCLUSION

At the completion of product development, product engineers should have qualified the product and the manufacturing processes for production. A final review should be conducted with the customer to assure satisfaction. A specification sheet should be generated which lists the device attributes and margin.

The product development efforts should be documented in a characteristics and development report. The report should include design specifics, process maps and manufacturing details, integration steps and post mortem analysis. Included at the end should be a section on lessons learned. This information is invaluable for future product development work.

Product development entails maturing the design and all the processes needed to produce the product. The work requires a dedicated team, careful planning, attention to detail and patience. The level of rigor is dependent on the device manufactured and the end use application. Product production depends strongly on good product development.

## 4.0 ACKNOWLEDGEMENTS

The authors thank the MEMS development engineers at Sandia National Laboratories and Honeywell FM&T, Kansas City, Missouri for their contributions to this work. Special thanks go out to Thomas Swiler, Carl Gustafson, Gregory Bogart Randy Shul, Andrew Oliver, Jeremy Walraven, Danelle Tanner, Fredd Rodriguez, Carol Sumpter and Julia Hammond for their contributions to MEMS product development.

## 5.0 REFERENCES

[1] M. Polosky, G Sleefe, "Electrical and Optical Integration of Microsystems", International Workshop on MEMS and Nanotechnology Integration (MNI): Applications, 10-11 May 2004, Montreux Switzerland

[2] M. Polosky et al, Trajectory Safety Subsystem on a Chip (TSSC)," Sensors and Actuators 1999, Sendai, Japan

[3] T. George, R. Powers, "Closing the TRL Gap," Aerospace America, pp. 24-6, August 2003.

[4] R.E. Shannon, "Engineering Management," Wiley, USA, 1980

[5] M.R. Daily, C.W. Sumpter, "A Structured Process for the Transitioning New Technology into Fieldable Products", IEEE International Engineering Management Conference, St Johns, Newfoundland, Canada, 2005

[6] G.R. Blackwell, "The Electronic Packaging Handbook," CRC Press, USA, 2000